\documentclass[conference]{IEEEtran}
\IEEEoverridecommandlockouts
\usepackage{cite}
\usepackage{amsmath,amssymb,amsfonts}
\usepackage{algorithmic}
\usepackage{graphicx}
\usepackage{textcomp}
\usepackage{xcolor}
\usepackage{url} 
\def\BibTeX{{\rm B\kern-.05em{\sc i\kern-.025em b}\kern-.08em
    T\kern-.1667em\lower.7ex\hbox{E}\kern-.125emX}}

\usepackage{fancyhdr}
\begin{document}

\title{Video-based Social Interaction Behavior Analysis with the Simulated Interaction Task for Children (Kids-SIT)
\thanks{This study was funded by the Federal Ministry of Education and Research Germany (BMBF; 01IS20046), the Deutsche Forschungsgemeinschaft (DFG; AS 553/1-1;), the Excellence Strategy of the Federal Government and the Länder by the Berlin University Alliance.}%
\thanks{Accepted at ACII 2026. © 2026 IEEE. Personal use of this material is permitted. Permission from IEEE must be obtained for all other uses, in any current or future media, including reprinting/republishing this material for advertising or promotional purposes, creating new collective works, for resale or redistribution to servers or lists, or reuse of any copyrighted component of this work in other works.}
    \thanks{$^{*}$These authors contributed equally to this work.}%
    \thanks{$^{\dagger}$Corresponding author, drimalla@techfak.uni-bielefeld.de}

}

\author{\parbox{\textwidth}{\centering
    {\large Rituja Pardhi$^{1*}$, Matthias Norden$^{1*}$, William Saakyan$^1$, Nadine Vietmeier$^2$, Simone Kirst$^2$, \\ Isabel Dziobek$^2$, Julia Asbrand$^3$, Hanna Drimalla$^{1\dagger}$}\\
    {\normalsize
    $^1$ Human-Centered Artificial Intelligence Group, Bielefeld University, Bielefeld, Germany\\
    $^2$ Institute of Psychology, Humboldt-Universität zu Berlin, Berlin, Germany\\
    $^3$ Institute of Psychology, Friedrich-Schiller-University of Jena, Jena, Germany}
    }}
    
\maketitle
\thispagestyle{fancy}

\begin{abstract}

Accurately quantifying children’s social interaction behavior is part of understanding their cognitive and emotional development, as well as mental health conditions. Kids-SIT is a web-based tool designed to computationally analyse children’s behaviors by engaging them in a standardized video conversation while their responses are video recorded. In a pre-registered study with 21 healthy children and 12 children diagnosed with social anxiety disorder (SAD), aged 9-14 years, we assess its potential as an accessible paradigm for automated analysis of children's social interaction behavior. 
We evaluate whether the Kids-SIT can elicit naturalistic interaction patterns in healthy children, and how well automatic feature extraction methods can detect these patterns. To this end, we analyse children’s subjective impressions, verbal responses, and non-verbal behaviors. Non-verbal behaviors were  manually annotated and, independently, automatically extracted using state-of-the-art methods. In an exploratory analysis, we further assess whether automatically extracted features can distinguish between children with and without SAD.
Verbal responses and post-hoc impressions indicate that the Kids-SIT elicits natural social interaction behavior. Non-verbal behavior aligned with this pattern: children looked at their interaction partner most of the time, particularly while listening rather than speaking. Smiling and gazing toward the partner occurred more frequently during the person-directed liked and disliked parts of the conversation than during the picture-description phase. These patterns were captured by both annotations and computational methods. Automatically extracted features enabled above-chance differentiation between children with and without SAD.
Our results underscore the potential of the Kids-SIT as a scalable tool for analysing children’s social interaction behavior, with applicability extending to clinical contexts.

\end{abstract}


\section{Introduction}

How children behave in social contexts conveys important information on their cognitive, emotional, and social development as well as on their mental health \cite{Gottman1975_}. Altered social interaction behavior is commonly observed in autism spectrum conditions (ASC), social anxiety disorder (SAD), attention-deficit/hyperactivity disorder (ADHD), and depression where atypical patterns of verbal and non-verbal communication can lead to challenges in families and schools \cite{Cresswell2019_Theexperiencesofpeerrelationshipsamongstautisticadolescents, darling2021behavioral}. Current clinical assessment methods of social interaction behavior often rely on subjective measures such as parent- and self-reports, clinician-administered interviews, and labor-intensive behavioral observation protocols \cite{dsm_v}. These methods are time-consuming, require trained personnel, and their subjectivity can yield inconclusive or inconsistent findings \cite{de2015introduction, de2015validity}.

Analysing verbal and non-verbal social interaction behaviors from video recordings using social signal processing and machine learning (ML) has shown promising results for various use cases, including ASC \cite{Saakyan2023_, saakyan2025improving}, depression, and SAD diagnosis \cite{Pampouchidou2020_Automatedfacialvideo-basedrecognitionofdepressionandanxietysymptomseverity}.
Despite this progress, using such methods for assessing children’s social interaction behavior remains a challenging task. Most machine learning studies focusing on children rely on complex, resource-intensive setups, often analysing caregiver-child \cite{Messinger2009_AutomatedMeasurementofFacialExpressioninInfant-MotherInteraction, Karaca2024_} or experimenter-child interactions \cite{Bertamini2021_QuantifyingtheChild-TherapistInteractioninASDIntervention}, and uncontrolled settings such as classrooms \cite{Messinger2022_}. Despite offering valuable insights, these approaches are difficult to standardize and scale, limiting their widespread and repeated use and integration into standard care practices.

The \emph{Simulated Interaction Task for Children (Kids-SIT)} \cite{ Norden2024_IntroducingtheSimulatedInteractionTaskforChildrenKids-SIT} 
offers a standardized, scalable, and accessible means of eliciting and assessing social interaction. %
The Kids-SIT is a web-based application that enables children to participate in a seven-minute structured conversation with a pre-recorded actress, alternating speaking and listening across a neutral picture-description phase and emotionally salient liked and disliked-food phases, while their interaction is video-recorded. The recorded data are then computationally analysed to extract and quantify social interaction behaviors. A demonstration version of the Kids-SIT prototype can be tested at \url{https://www.simulatedinteraction.com/children_1}.

First, we evaluate whether children actively engage with the Kids-SIT to ensure that the paradigm elicits naturalistic social interaction. We consider both verbal and non-verbal markers of interaction. To assess verbal markers, we transcribe children’s speech and examine their responsiveness and alignment with conversational topics. 
For non-verbal interaction markers, we focus on three core behaviors - gaze (as a proxy for engagement \cite{vertegaal2001eye}), smiling (reflecting emotional responsiveness \cite{niedenthal2010simulation}), and nodding (as a backchannel cue \cite{aburumman2022nonverbal}) - selected for their well-established relevance in social interaction research. Together, these non-verbal measures allow us to determine whether children are attentive and responsive during the interaction, which we assess using human-annotated data.

Second, we focus on the automated analysis of children’s behavior, an approach increasingly used to study social interaction \cite{eschman2022remote, crescenzi2025facial}. Most widely used algorithms such as OpenFace 2.0 \cite{Baltrusaitis2018_OpenFace2.0}, are trained and validated on adult datasets, raising concerns about their generalizability to children, with prior work documenting poor performance \cite{onal2023infant, dapogny2019automatically}. To address this limitation, we apply and compare three standard automated behavioral feature extraction techniques in terms of their alignment with human annotations: OpenFace 2.0 \cite{Baltrusaitis2018_OpenFace2.0} for smile, gaze angles, and head pose, PyAFAR \cite{hindujapyafar,ertugrulpyafar} for smile and head pose, and Rehg Lab’s eye-contact-cnn \cite{chong2020} for gaze, the latter two designed specifically for children. We examine whether these tools can capture naturalistic behavioral patterns by assessing their alignment with human annotations, particularly for off-screen gaze deviations, smiles, and nods.

In the above two steps, we focus on validating the Kids-SIT using behavioral data from healthy children. The ultimate goal of Kids-SIT is its application in clinical contexts, particularly for the assessment of social interaction difficulties associated with conditions such as ASC, SAD, and ADHD. Thus, we conduct a preliminary investigation using a machine learning approach to determine whether the paradigm elicits non-verbal behaviors that are discriminative between children with SAD and those without, and whether these differences can be detected using automatically extracted behavioral features.

The major contributions of our work are:
\begin{itemize}

\item We deploy the Kids-SIT and validate, via manual annotation, that it elicits naturalistic behavioral patterns aligned with prior literature.

\item We develop an automated behavioral analysis pipeline by extracting gaze, smile, and head pose features using state-of-the-art methods, introducing a rule-based and data-driven nod detection algorithm, and determining empirically validated thresholds that show improved alignment with human annotations.

\item We analyse the alignment between manual annotations and computationally extracted behaviors, highlighting both convergences and discrepancies, and demonstrate that these features are informative of relevant social interaction patterns in children.

\item We apply an exploratory machine-learning-based classification to distinguish between 21 healthy children and 12 children with SAD, indicating the potential of the Kids-SIT for future clinical applications.
\end{itemize}

\section{RELATED WORK}

\subsection{Assessing Children's Social Interaction Behavior}

Children’s social interaction behavior is commonly assessed through structured observational instruments that capture eye contact, gestures, and turn-taking, such as the Early Social-Communication Scales \cite{Steiner2013_} and the Autism Diagnostic Observation Schedule (ADOS \cite{Lord1989_Autismdiagnosticobservationschedule}). While these provide detailed and clinically valid information, they are time-intensive, require coding expertise, and might differ across experimenters. To enable more objective quantification, researchers increasingly use technology-based solutions: %
eye-tracking has been used to investigate visual attention, joint attention, and gaze-following behaviors %
\cite{Tonsing2025_Alteredinteractivedynamicsofgazebehaviorduringface-to-faceinteractioninautisticindividuals, Ambarchi2024_Socialandjointattentionduringsharedbookreadinginyoungautisticchildren, Yoon2025_CorrelationBetweenGazeBehaviorsandSocialCommunicationSkillsofYoungAutisticChildren}, and virtual reality (VR) to elicit and study gaze dynamics in controlled but ecologically valid contexts \cite{Parsons2017_, Elkin2022_GazeFixationandVisualSearchingBehaviorsduringanImmersiveVirtualRealitySocialSkillsTrainingExperienceforChildrenandYouthwithAutismSpectrumDisorder}.
However, these approaches require specialized equipment, trained personnel, and controlled settings, limiting their scalability. 

\subsection{Video-mediated Social Interaction Paradigms}
Structured video-mediated paradigms offer a %
standardized and scalable approach to assessing social behavior. Children with ASC showed acceptance of face-to-face, video call, and pre-recorded video interactions \cite{Pliska2023_}, suggesting digital platforms can effectively replicate real-world social interactions for this population, and Howell et al. \cite{Howell2024_Mutualeyegazeandvocalpitchinrelationtosocialanxietyanddepression} linked mutual gaze and paralinguistic features to SAD and depression in young adults using a video-call-like task in VR.
The original ``Simulated Interaction Task" (SIT) \cite{Drimalla2020_Towardstheautomaticdetectionofsocialbiomarkersinautismspectrumdisorder} was developed to objectively assess social interaction behaviors in adults, particularly for ASC diagnosis, achieving performance comparable to clinical expert ratings using automated analysis of facial, vocal, and gaze features \cite{Drimalla2020_Towardstheautomaticdetectionofsocialbiomarkersinautismspectrumdisorder, Saakyan2023_, saakyan2025improving}. Norden et al. \cite{Norden2024_IntroducingtheSimulatedInteractionTaskforChildrenKids-SIT} 
adapted this paradigm for children (Kids-SIT), but it has not yet been empirically applied or validated.
In addition, the feasibility of employing computational behavioral feature extraction systems in this setting, as previously demonstrated in adult studies, has yet to be systematically evaluated.

\subsection{Video-based Computational Analysis of Children’s Interaction Behaviors}
While smiles, gaze deviation, and nodding have been extensively extracted from adult videos %
\cite{Baltrusaitis2018_OpenFace2.0, cheong2023py}, few algorithms target children.
OpenFace 2.0 \cite{Baltrusaitis2018_OpenFace2.0} is widely used for automated feature extraction, including in the adult SIT \cite{Drimalla2020_Towardstheautomaticdetectionofsocialbiomarkersinautismspectrumdisorder, Saakyan2023_}. %
Dapogny et al. \cite{dapogny2019automatically} emphasized that models for facial expression analysis trained on adult datasets perform poorly when applied to children. Several child-focused approaches address this: %
HARMONI \cite{Weng2025_} applies a 3D CNN to quantify caregiver–child interactions from observational videos, but while it captures key interaction dynamics, it does not explicitly model core social cues such as smile and gaze dynamics. Guo et al. \cite{Guo2023} developed a mutual-gaze framework whose ML regression captured human-coded patterns.
PyAFAR \cite{ertugrulpyafar, hindujapyafar} performs facial action unit (AU) detection in infants and adults, outperforming OpenFace 2.0 on the GFT dataset.
Rehg’s eye-contact-cnn \cite{chong2020} is a deep learning-based model specifically trained to detect eye contact in children, achieving human-comparable performance on a diverse ASC/non-ASC sample.

\section{METHODS}

To enhance transparency, all hypotheses and analysis plans were pre-registered a priori on OSF (\url{https://osf.io/tw9ne/}) based on existing literature.
The code for processing the videos and analysing the behaviors is available on GitHub (\url{https://github.com/mbp-lab/kids-sit}). 

\subsection{Data Collection}
We recruited 21 healthy children aged between 9 and 14 years (11 female, 10 male; $M_{\text{Age}} = 11.6$ years) and 12 children (9 female, 3 male; $M_{\text{Age}} = 11.9$ years) diagnosed with SAD. Ethical approval was granted by the ethics committee of Humboldt University Berlin (application 2021-26%
). For details on the study procedure and psychological assessment see supplementary material (section I).

\subsection{Subjective impression}

To measure children's impressions of the Kids-SIT, short 7-point Likert-scale questions were displayed after the interaction. Children were asked how much they liked/disliked the meals that the actress talked about; how uncomfortable they felt due to the actress being pre-recorded and how similarly they behaved during the Kids-SIT compared to a normal conversation. Children were also asked to rate situational anxiety and arousal on a Likert scale ranging from 0 (``not at all") to 10 (``very much") before and after Kids-SIT.

\subsection{Verbal Responses}

To examine verbal interaction, we focused on children’s speech, evaluating responsiveness and topical alignment.
According to our pre-registered hypothesis, we expected at least one sentence per topic (i.e., picture, liked, and disliked food) and that responses match the topics in their word choice.
Video recordings were automatically transcribed using the Whisper large speech-to-text model \cite{Radford2023_} and subsequently manually corrected. The transcriptions were then analysed using the linguistic inquiry and word count (LIWC) \cite{boyd2022liwc22}, a dictionary-based text-analysis tool that computes the proportion of words belonging to predefined categories, with a German dictionary \cite{Meier2019_LIWCaufDeutsch}. Extracted measures included word counts and the fractions of the word categories “ingest” (related to eating or consuming), “posemo” (positive emotions like joy or love), “negemo” (negative emotions like sadness or anger), and “see” (words related to vision or perception).
Word counts and fractions of word categories across conversation parts were evaluated using repeated measures Friedman test after checking for normality. Bonferroni corrected pairwise post-hoc tests are reported.

\subsection{Non-verbal Responses}

To examine whether children's non-verbal behavior resembled naturalistic social interaction as observed in previous literature, and to assess how well this behavior could be extracted computationally, we analysed gaze, smiling, and nodding in video recordings. We only analysed videos of the conversation parts considering the bilateral descriptions of a busy scene, liked and disliked foods. We chose this approach in order to compare children's social interaction behavior across listening and speaking as well as across emotionally differently toned parts. 

This analysis was performed using manual annotations (see \ref{sec:annotations}) and three different computational methods - OpenFace 2.0 \cite{Baltrusaitis2018_OpenFace2.0}, PyAFAR \cite{hindujapyafar,ertugrulpyafar}, and eye-contact-cnn \cite{chong2020}. 
PyAFAR and eye-contact-cnn were evaluated because their training data closely match the child population examined here, hence supposedly leading to more accurate behavior extractions than OpenFace which is trained on adult datasets. 

\vspace{1em}
\subsubsection{Gaze}

\textbf{Hypothesis:} Based on the study by \cite{Grossman2019_Facetimevs.Screentime}, which reports that healthy children fixate on their interaction partner’s face for 29\% of the time during live interactions, we adopted a 25\% expectation for gaze toward the screen (i.e., \textless\ 75\% gaze deviation) as a conservative estimate that accounts for attention distributed between the face and the screen. \textbf{Manual annotation:} To evaluate the hypothesis, we first analysed the human annotations of off-screen gaze deviation (see section \ref{sec:annotations} for annotation details) and used a one-sample Wilcoxon signed-rank test to check whether it exceeded 75\% of the overall time. \textbf{Automated extraction:} To investigate how well this behavior pattern could be extracted computationally, we applied \textbf{\textit{OpenFace}} \cite{Baltrusaitis2018_OpenFace2.0} to extract gaze angles. Based on these, we used an off-screen gaze deviation detection algorithm
\cite{saakyan2025improving}:
the gaze angles were mapped to positions relative to the screen using estimations of camera- and eye-to-screen distances and the screen resolution. Additionally, we employed \textbf{\textit{eye-contact-cnn}} library \cite{chong2020} to obtain an eye-camera contact score between 0 and 1. %
These scores were binarized and aggregated across frames to calculate the percentage of time children deviated their gaze from the screen.

\vspace{1em}
\subsubsection{Smile}

\textbf{Hypothesis:} Following the findings of \cite{Drimalla2020_Towardstheautomaticdetectionofsocialbiomarkersinautismspectrumdisorder},
we hypothesized that children would smile more frequently during the liked and disliked food parts than in the picture description part. \textbf{Manual annotation:} To assess this hypothesis, we first evaluated the human annotations of smiling behaviors. %
After testing for normality, paired t-tests or Wilcoxon signed-rank tests were applied as appropriate to compare percentages of smiling activity between different parts. \textbf{Automated extraction:} To determine how effectively this pattern could be detected automatically, we applied \textbf{\textit{OpenFace}} \cite{Baltrusaitis2018_OpenFace2.0} to extract action unit (AU) occurrences in each frame. We used the binary occurrence feature of AU12 (Lip Corner Puller) to detect smiles.
Following this, we also employed \textbf{\textit{PyAFAR’s}} infant model \cite{hindujapyafar,ertugrulpyafar}, to extract AU occurrence likelihoods. %
These likelihoods were thresholded to obtain a binary indicator of smiling, and frame-level outputs were combined to estimate the proportion of time smiling occurred within each segment.

\vspace{1em}
\subsubsection{Nod}

\textbf{Hypothesis:} Previous work indicates that children, like adults, produce backchannel behaviors such as head nods \cite{liu2022predicting, bodur2023using}. We therefore hypothesized more nodding during listening than speaking. \textbf{Manual annotation:} %
Nods were annotated by three independent raters using majority vote. Differences between listening and speaking were tested using paired t-tests or Wilcoxon signed-rank tests after normality checks. \textbf{Automated extraction:} To assess automatic detection, we developed a rule-based nod detection algorithm using pitch and yaw head pose estimates extracted using \textbf{\textit{OpenFace}} and the \textbf{\textit{PyAFAR}}-infant model. A detailed description of the nod detection algorithm, together with the selected threshold values, is provided in the supplementary material (section II). 
The algorithm produced a binary frame-level nod indicator, used to compute the proportion of nodding time within each part. It was applied independently to head pose estimates obtained from OpenFace and PyAFAR, using library-specific threshold settings to account for differences in noise characteristics. %

\subsection{Video Annotations of Non-Verbal Responses} \label{sec:annotations}

Annotators with a university degree in psychology independently rated all Kids-SIT video recordings across six conversation parts using BORIS \cite{Friard2016_BORISa}. Three non-verbal behaviors were annotated: smile, gaze, and nod. Smile and gaze were annotated by two annotators, and nod by three annotators. We evaluate inter-annotator agreement at instance level and frame level. For statistical tests, we used only frames or instances for which annotators were in agreement. For more details on annotation procedure and inter-annotator agreement, refer to supplementary material (section III).

\subsection{Threshold selection} \label{sec:threshold}
Using default thresholds for the detection of smiles and gaze behavior (e.g., 0.5) might lead to over- or undersensitivity and hence misleading interpretations of the results. We systematically tested threshold values ranging from 0.1 to 0.9 (in steps of 0.1). For each threshold, binary labels were generated and agreement with two annotators was computed using Cohen’s kappa. We report the optimal thresholds and annotator agreements for each algorithm and use these for comparing computationally found smile and off-screen gaze deviation across Kids-SIT parts.

\subsection{Kids-SIT in Clinical Context} 

As a step toward the clinical applicability, an initial assessment was conducted using binary classification based on automatically extracted behavioral features to distinguish between children with SAD and those without.
Head pose and facial features extracted using PyAFAR-infant, and gaze features extracted using eye-contact-cnn were used. Participant-level features were derived from frame-wise behavioral signals, including head pose estimates (pitch, yaw, and roll), facial action unit occurrences (AU4, AU6, AU12), and gaze behavior. For each participant, mean and standard deviation statistics were computed for each feature across the interaction to obtain participant-level representations. A gradient-boosted decision tree model (XGBoost \cite{chen2016xgboost}) was trained on these aggregated features. Model evaluation was performed using leave-one-participant-out cross-validation and classification performance was quantified using the area under the receiver operating characteristic curve (AUC).

\section{RESULTS}

To validate the Kids-SIT, we first analysed data from healthy children (Sections A–C). Section D presents results for children with SAD and the exploratory clinical applicability.

\subsection{Subjective impression}

According to their responses to the 7-point Likert scales, children indicated that they did not feel uncomfortable due to the actress being pre-recorded (\textit{Mdn} = 1, \textit{IQR} = 0–1) and behaved naturally during the test (\textit{Mdn} = 5, \textit{IQR} = 4–6). Most children liked the favorite food of the actress (\textit{Mdn} = 5, \textit{IQR} = 3–6) while disliking the example for disgust food (\textit{Mdn} = 0, \textit{IQR} = 0–1). There was no evidence for significant changes in anxiety and arousal from pre to post Kids-SIT (both \textit{p} \textgreater\ 0.05). 

\subsection{Verbal Responses}

\begin{figure*}[tb] %
    \centering
    \includegraphics[width= 0.6 \textwidth]{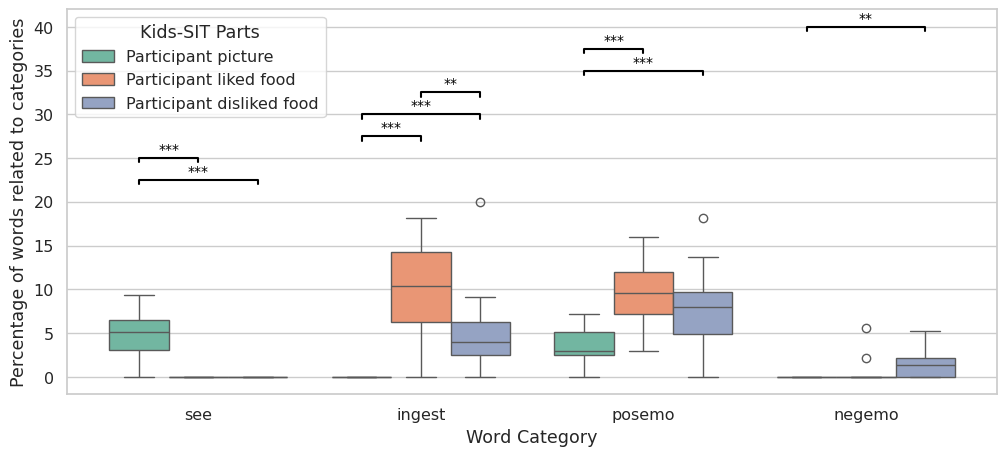}
    \caption{Percentages of words related to seeing something (see), ingestion and food (ingest), positive emotions (posemo) and negative emotions (negemo) across children's verbal responses to respective Kids-SIT parts. Significant differences between categories are annotated according to Bonferroni corrected pair-wise Wilcoxon-signed rank tests with significance levels of 0.01 (**) and 0.001 (***).}
    \label{fig:LIWC_plot}
\end{figure*}

The minimum word count across all response parts was 14, while the median length of responses across parts ranged from 34 (liked food) to 47 words (disliked food). Fractions of word categories according to LIWC and statistical comparisons across the Kids-SIT parts are presented in Fig. \ref{fig:LIWC_plot}. Responses significantly differed across parts for the word categories related to seeing (\textit{F}(2, 40) = 78.74, \textit{p} \textless\ 0.001), positive emotions (\textit{F}(2, 40) = 20.95, \textit{p} \textless\  0.001), negative emotions (\textit{F}(2, 40) = 20.95, \textit{p} \textless\  0.001), and food or ingestion (\textit{Q}(2) = 32.70, \textit{p} \textless\ 0.001). Responses included significantly more words related to seeing and describing for the picture description (\textit{Mdn} = 5.13\%) than for the liked (\textit{Mdn} = 0\%) and disliked food (\textit{Mdn} = 0\%) (all \textit{p} \textless\ 0.001). Similarly, both food related responses included significantly more words related to food and ingestion than the picture description (\textit{Mdn} = 0\%) response (all \textit{p} \textless\ 0.001) while the response on liked food (\textit{M} = 10.34\%) included significantly more respective words than the disliked food response (\textit{M} = 5.13\%, \textit{p} \textless\  0.01). Words related to positive emotions appeared significantly more frequently in the liked (\textit{Mdn} = 9.52\%) and disliked food response (\textit{Mdn} = 8.00\%) than during the picture description (\textit{Mdn} = 2.94\%, all \textit{p} \textless\  0.001). Words related to negative emotions appeared significantly more frequently in the disliked food response (\textit{Mdn} = 1.3\%) than in the picture description (\textit{Mdn} = 0\%, \textit{p} \textless\  0.01).

\subsection{Non-verbal Responses}

\begin{figure}[tb] %
    \centering
    \includegraphics[width=\linewidth]{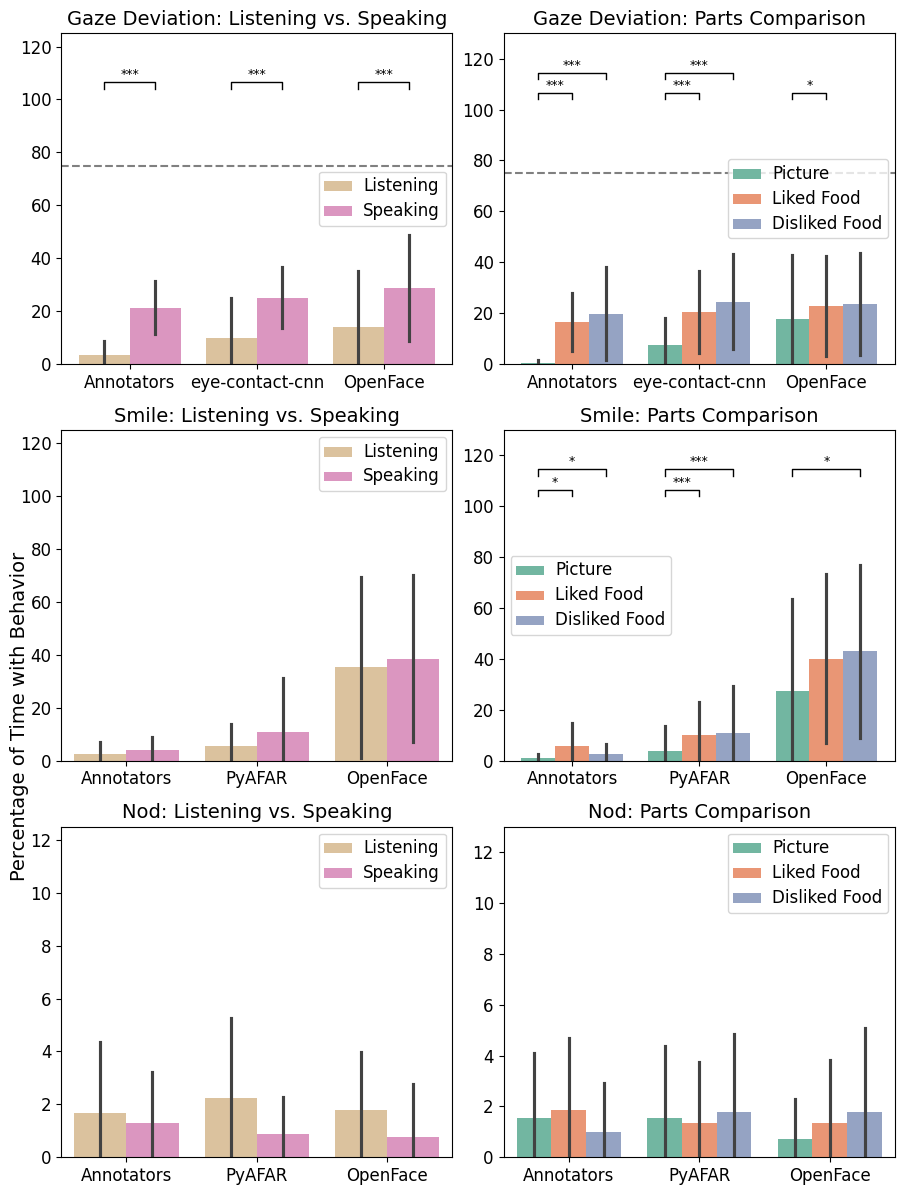}
    \caption{Percentages of gaze, smiling and nodding behaviors across speaking vs. listening  and conversation topics. Error bars represent standard deviations. Dashed line in the first two plots represents literature based expected maximum gaze deviation threshold \cite{Grossman2019_Facetimevs.Screentime}. Significance levels are indicated as 0.05 (*), 0.01 (**), and 0.001 (***).}
    \label{fig:behavior_annotation_plot}
\end{figure}

Counts of frames and instances for gaze deviation, smiling, and nodding (annotators and computational methods) can be found in supplementary material (section IV). 
The distribution of percentages of time for the three behaviors within the different Kids-SIT parts can be seen in Fig. \ref{fig:behavior_annotation_plot}. 

\vspace{1em}
\subsubsection{Gaze}

\textbf{Manual annotation:} Inter-rater instance level agreement for off-screen gaze deviation was 94\%, with frame level Cohen’s Kappa coefficient of 0.84. According to their manually annotated off-screen gaze behavior, children gazed away from the screen for 12\% over all six interaction parts which was significantly lower than 75\% (\textit{t} = 8.1, \textit{p} \textless\ 0.0001) as described in previous work \cite{Grossman2019_Facetimevs.Screentime}.
They gazed significantly more off the screen when they spoke (\textit{Mdn} = 21.06\%) than when they listened (\textit{Mdn} = 1.26\%), \textit{t} = 8.5, \textit{p} \textless\ 0.0001. Children also gazed away for more of the time during liked (\textit{Mdn} = 14.73\%) and disliked (\textit{Mdn} = 16.75\%) food parts than during the picture description part (\textit{Mdn} = 0\%), both \textit{p} \textless\ 0.001.

\textbf{Automated extraction:} According to OpenFace, children gazed off-screen for a median of 16.31\% of the time and 14.26\% according to eye-contact-cnn. The findings were significantly lower than the hypothesized 75\% for both methods, OpenFace (\textit{W} = 3.0, \textit{p} \textless\ 0.0001) and eye-contact-cnn (\textit{W} = 0.0, \textit{p} \textless\ 0.0001). Children gazed significantly more off-screen during the speaking than during listening across both OpenFace (\textit{Mdn} = 28.44\% vs. \textit{Mdn} = 5.71\%), \textit{t(20)} = 4.09, \textit{p} \textless\ 0.001 and eye-contact-cnn (\textit{Mdn} = 23.90\% vs. \textit{Mdn} = 6.39\%), \textit{W(20)} = 14, \textit{p} \textless\ 0.001. Across both methods, gaze deviations were lower for the picture description part (\textit{Mdn} = 9.03\% for OpenFace; \textit{Mdn} = 3.44\% for eye-contact-cnn) than the liked food (\textit{Mdn} = 22.08\%, \textit{W} = 55, \textit{p} \textless\ 0.05 for OpenFace; \textit{Mdn} = 15.44\%, \textit{t} = 3.8, \textit{p} = 0.001 for eye-contact-cnn) and disliked food (\textit{Mdn} = 22.30\%, \textit{W} = 49, \textit{p} = 0.06 for OpenFace; \textit{Mdn} = 17.71\%, \textit{W} = 23, \textit{p} \textless\ 0.001 for eye-contact-cnn) parts.

\begin{figure}[b] %
    \centering
    \includegraphics[width=0.9\linewidth]{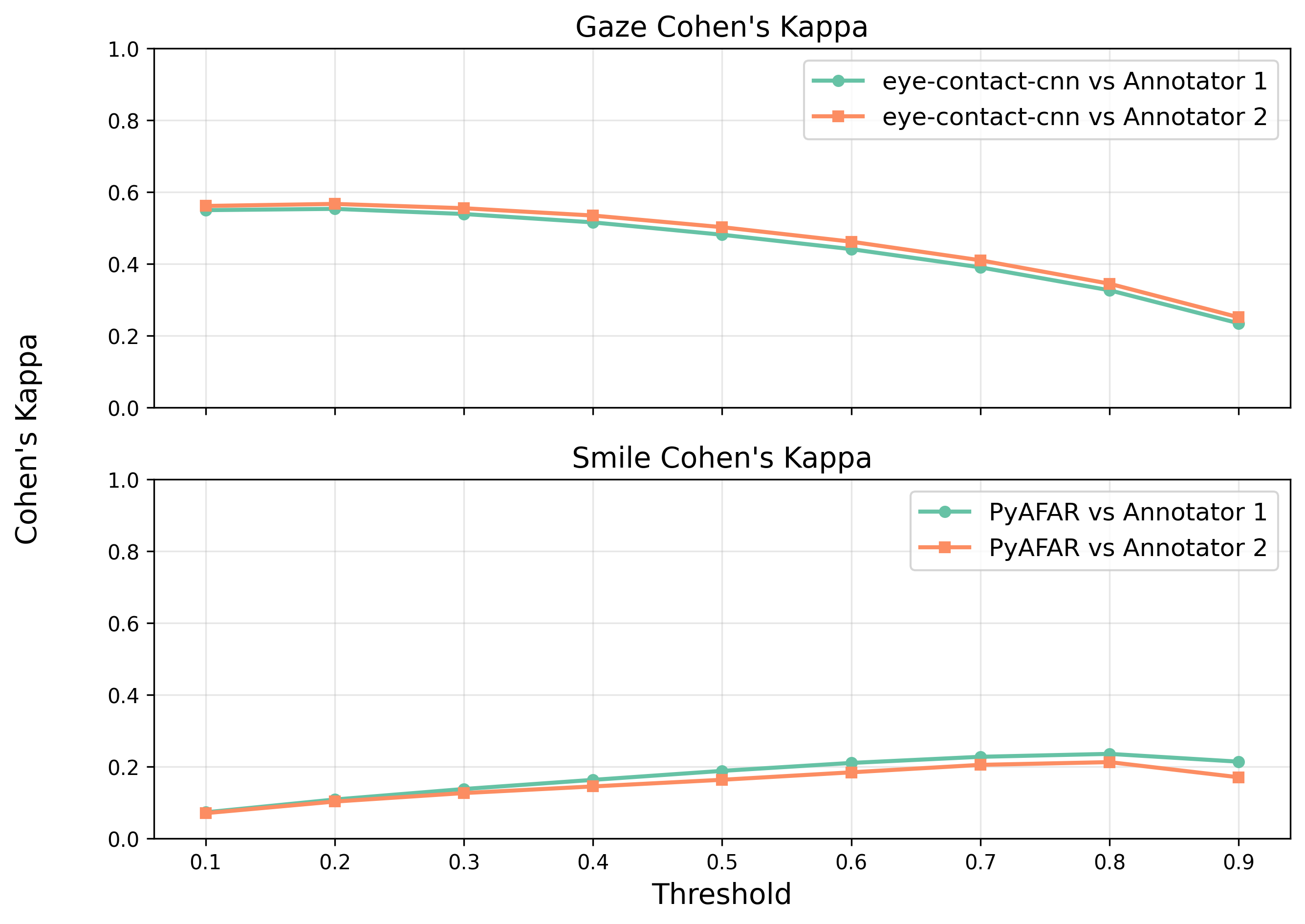}
    \caption{Sensitivity analysis of model–annotator agreement with respect to the threshold for gaze deviation and smile detection.
    }
    \label{fig:threshold_plot}
\end{figure}

\textbf{Manual vs Automated:} Computational methods yielded significantly more gaze deviation instances than annotators (all \textit{p} \textless\ 0.001), with mean instance counts per child being more than four times higher.
Frame level agreement between eye-contact-cnn and annotations across thresholds is shown in Fig. \ref{fig:threshold_plot} (top); a threshold of 0.2 yielded the highest agreement for eye-contact-cnn and was selected as optimal. Instance-level agreement with annotators was 38\% and 42\% for OpenFace, with corresponding frame-level Cohen’s kappa of 0.28 and 0.29. For eye-contact-cnn, instance-level agreement was 35\% and 38\%, with kappa values of 0.55 and 0.56.

\vspace{1em}
\subsubsection{Smile}

\textbf{Manual annotation:} Inter-rater instance level agreement for smile occurrence was 88\%, with a Cohen’s Kappa coefficient of 0.46. According to their annotated behaviors, children smiled for less of the time during the picture description part (\textit{M} = 0.91\%) when compared to the liked food (\textit{M} = 5.90\%), \textit{W} = 11, \textit{p} \textless\ 0.05 and disliked food part (\textit{M} = 2.73\%), \textit{W} = 7, \textit{p} \textless\ 0.05.

\textbf{Automated extraction:} There was no evidence for significant differences of smiling (i.e., AU12) behavior between speaking and listening parts for both annotated and computationally extracted behaviors. The analysis for part comparison for OpenFace showed that children \textit{smiled} for less of the time during the picture description part (\textit{Mdn} = 5.67\%) when compared to the liked food (\textit{Mdn} = 46.23\%), \textit{W} = 43, \textit{p} = 0.06 and disliked food part (\textit{Mdn} = 49.46\%), \textit{t} = 2.6, \textit{p} \textless\ 0.05. Similarly, PyAFAR detected less smiling during the picture description part (\textit{Mdn} = 0.2\%) compared to the liked food (\textit{Mdn} = 3.11\%), \textit{W} = 1, \textit{p} \textless\ 0.001 and disliked food part (\textit{Mdn} = 4.08\%), \textit{W} = 6, \textit{p} \textless\ 0.001.

\textbf{Manual vs Automated:} The number of smile instances extracted by OpenFace (\textit{M} = 28) was significantly higher than the mean of both annotators (all \textit{p} \textless\ 0.001) with mean instance counts per child being more than eight times higher. The instance counts detected by PyAFAR (\textit{M} = 100) were substantially higher than those recorded by annotators (both \textit{M} = 3.3). 

Fig. \ref{fig:threshold_plot} (bottom) shows the frame-level agreement between PyAFAR and annotations across thresholds. An optimal threshold of 0.8 was selected for smile detection using PyAFAR. The instance level agreement between OpenFace and the annotators was 22\% for both annotators. The associated Cohen’s Kappa values, computed at the frame level, were 0.14 and 0.11. PyAFAR achieved instance level agreements of 7\% and 6\% for annotators 1 and 2, respectively, with corresponding frame level Cohen’s Kappa values of 0.23 and 0.21. In supplementary material, %
we show two illustrative examples of frame-wise annotations and extracted behaviors for default values and optimized thresholds (section V).

\vspace{1em}
\subsubsection{Nod}
The instance-level inter-rater agreement for nod occurrence was 78\%. Based on human annotations as well as OpenFace and PyAFAR based estimates, there was no significant difference in nodding rates between listening and speaking. The instance-level agreement for nod detection between OpenFace and the annotator majority vote was 59\%, whereas PyAFAR achieved an agreement of 62\%.

\subsection{Kids-SIT in Clinical Context}

\begin{figure}[tb] %
    \centering
    \includegraphics[width=\linewidth]{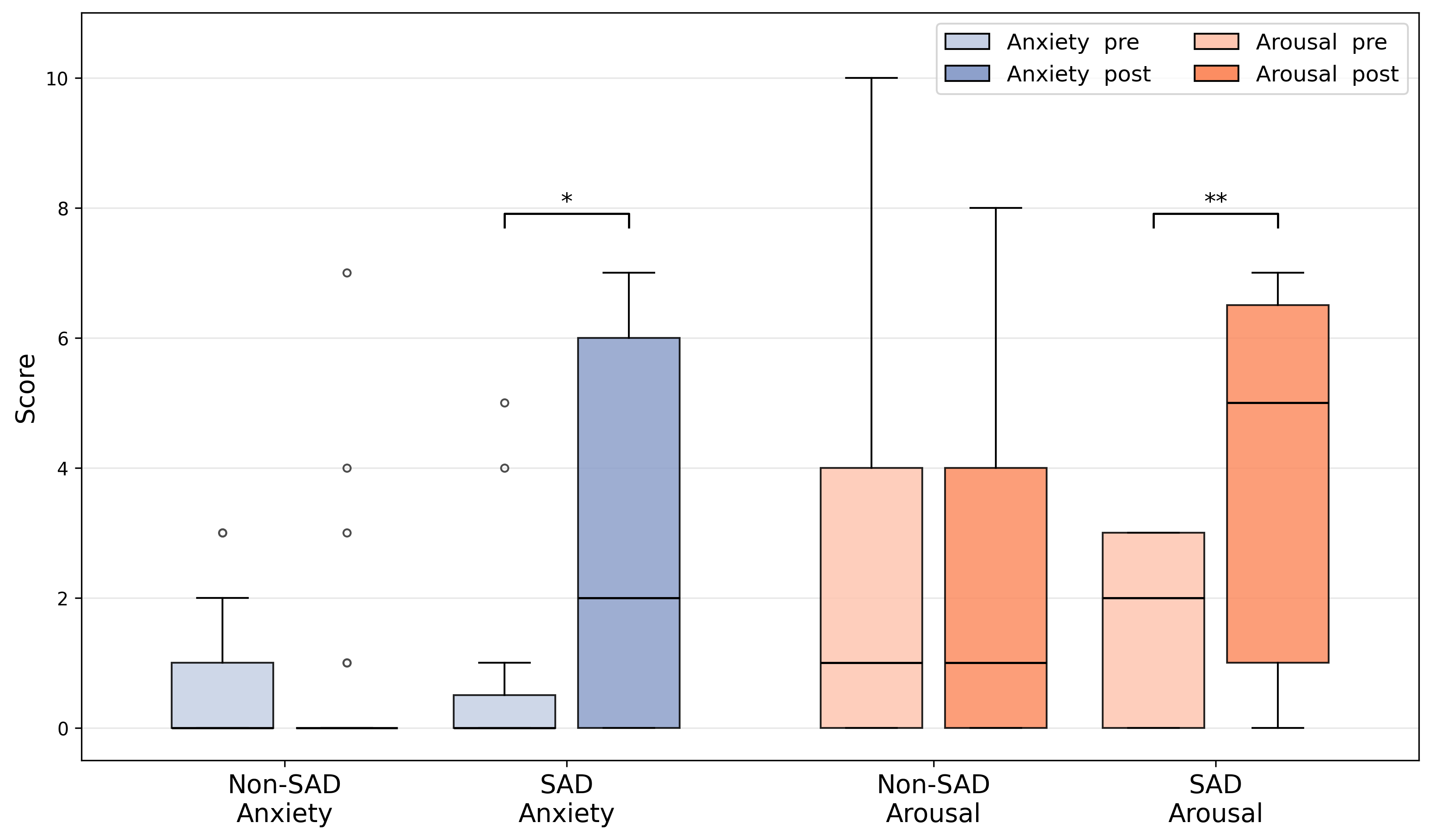}
    \caption{Children's responses on the 10-point situational anxiety and arousal questionnaires. Pre- and post-task responses are shown for each affect and group. Significant pre-to-post increases within each group are marked: * denoting \textit{p} \textless\ .05, ** denoting \textit{p} \textless\ .01.}
    \label{fig:pre_post}
\end{figure}

Similarly to the healthy group, children with SAD  generally liked the actress’s favorite food (\textit{Mdn} = 5, \textit{IQR} = 2.0–5.2), disliked the food she referred to as disgusting (\textit{Mdn} = 0, \textit{IQR} = 0.0–0.0), reported little discomfort with the pre-recorded actress (\textit{Mdn} = 1, \textit{IQR} = 0.8–2.2), and indicated that they behaved naturally during the task (\textit{Mdn} = 4, \textit{IQR} = 3.0–5.0). In contrast to the healthy group, children with SAD (n = 11; one participant excluded due to missing value) %
reported a significant increase in anxiety from pre (\textit{Mdn} = 0) to post (\textit{Mdn} = 2) Kids-SIT (\textit{W} = 0, \textit{p} \textless\ .05). Similarly, they reported a significant increase in arousal from pre (\textit{Mdn} = 2) to post (\textit{Mdn} = 5) Kids-SIT (\textit{W} = 2.5, \textit{p} \textless\ .01), as shown in Fig. \ref{fig:pre_post}.

Based on raw head pose, facial action unit, and gaze features automatically extracted by PyAFAR and eye-contact-cnn, the classification model achieved an average AUC of 0.74 under leave-one-participant-out cross-validation. Performance was stable across multiple random seeds (\textit{SD} = 0.015, \textit{range} = 0.71 - 0.76). In contrast, using engineered behavioral features (e.g., binary nod, smile, and gaze measures) resulted in reduced classification performance; results are reported in the supplementary material (section VI).

\section{DISCUSSION}

\subsection{Kids-SIT as a Social Interaction Paradigm}

Based on the children's feedback, they behaved naturally and did not feel uncomfortable due to the ``simulated" nature of the conversation. All children followed the conversation and responded to each topic, with LIWC confirming semantic alignment.
According to their annotated non-verbal behaviors, off-screen gaze was substantially below the 75\% expectation \cite{Grossman2019_Facetimevs.Screentime}. Consistent with \cite{Drimalla2020_Towardstheautomaticdetectionofsocialbiomarkersinautismspectrumdisorder}, smiling was higher in liked and disliked food parts than in the neutral picture description part. Although nodding was expected to be higher during listening than speaking, no significant difference was observed. To our knowledge, no prior work has directly compared nodding across pre-recorded and live interaction; it is therefore plausible that the pre-recorded format contributed to the reduced nodding, though this remains an open question for future research. Children nonetheless reported behaving naturally during the interaction, suggesting engagement was largely preserved.

This study provides initial insights into children's interaction behavior, though capturing subtle differences may require larger samples. 
Nevertheless, the alignment of annotated behavior patterns with prior literature supports the feasibility of using Kids-SIT to elicit naturalistic social interaction behaviors in children.

\subsection{Objective Automated Behavior Analysis}

Consistent with the manual annotation, gaze features extracted from OpenFace and eye-contact-cnn indicated that children looked at the actress most of the time and deviated far less than the hypothesized 75\% \cite{Grossman2019_Facetimevs.Screentime}, with greater deviation during speaking than listening. In line with the literature and annotations, frequency of smiles extracted by OpenFace and PyAFAR was higher in emotionally salient segments than in neutral segments. For nodding, neither method showed statistically significant difference between listening and speaking.

OpenFace detected more smile frames than PyAFAR and annotators. While PyAFAR detected substantially more smile instances than annotators, the overall number of smiling frames was similar, indicating that PyAFAR identifies many brief smile events, whereas human annotators mark longer occurrences. The automated tools yielded more frames and instances than manual annotation across behaviors, and such differences in absolute levels warrant closer analysis before these tools are used to estimate absolute behavioral rates.

Agreement between computational methods and annotators was lower than inter-annotator agreement. The lower agreement for OpenFace likely reflects its training on adult data, as prior work shows reduced performance on mismatched populations and dynamic video contexts\cite{Stockli2018_FacialexpressionanalysiswithAFFDEXandFACET}. 
In contrast, PyAFAR, trained on children’s data, showed higher agreement at the frame level for gaze and smiles and at the instance level for nods. The overall agreement discrepancy might also reflect higher sensitivity of algorithms for behavior recognition with subtle signals, potentially relevant for analysing intra- and interindividual differences. %

This finding suggests that OpenFace, PyAFAR, and eye-contact-cnn can feasibly capture children's naturalistic social interaction behaviors, specifically gaze and smile.

Selecting appropriate thresholds to convert continuous smile and gaze deviation likelihood scores into binary labels is a key challenge. %
Our systematic fine-tuning ensures that the automated detection of smiling and gaze deviation aligns more closely with human perception of these behaviors. Accordingly, the achieved results can be interpreted as an empirical upper bound under calibrated conditions; without such calibration, performance would likely deteriorate. %
We report the thresholds and agreement scores as reference values for future studies with Kids-SIT and related datasets, supporting reproducibility and enabling fairer comparisons.

\subsection{Kids-SIT in Clinical Context}
Children with SAD reported behaving naturally and experiencing little discomfort despite the “simulated” interaction. Compared to the healthy group, significant increases in anxiety and arousal were observed from pre to post Kids-SIT.

The classification results indicate that the Kids-SIT paradigm elicits non-verbal behavioral signals that can be captured using automated feature extraction methods and are informative for group-level differentiation. Due to the small sample size, this analysis is a first step and non-diagnostic, but it motivates future work with larger clinical cohorts. 

Building on these findings, Kids-SIT may be applied to other clinical conditions associated with atypical social interaction. Atypical gaze patterns are established markers for ASC, SAD, and ADHD \cite{mundy1986defining,fu2019stationary,miller2006right}, while reduced smiling has been linked to schizophrenia and depression \cite{parisi2025emotional,tremeau2005facial,lacerda2024high}. Applications range from condition classification to analysing interaction differences across and within disorders.

\section{CONCLUSION AND FUTURE WORK}

Kids-SIT elicited naturalistic social interaction in children while maintaining standardization and scalability. Automated methods captured key behavioral patterns consistent with annotations and prior work,  supporting their feasibility in extracting markers such as gaze and smile. Classification results provide initial evidence that Kids-SIT elicits behaviors that differentiate between children with and without SAD and that automated methods trained on child-specific data can effectively capture these signals.

Future work will extend Kids-SIT to a multi-center outpatient clinical study and establish a dataset of children's social interaction behavior with therapist-provided clinical information. Building on this, machine learning will be applied to larger clinical cohorts to assess robustness and generalizability and extend the approach to other conditions (e.g., ASC and depression). %
Efforts will also focus on validating ecological validity and expanding the set of analysed signals to include head and body movements.

\section*{ETHICAL IMPACT STATEMENT}

The deployment of a fully automated and accessible tool with the ability to collect video data and analyse sensitive behavioral indicators - including facial expressions, gaze direction, and verbal responses - in children, such as the Kids-SIT, warrants careful ethical consideration. In the following, we discuss the implications of using the Kids-SIT for assessing children’s social interaction behavior, outline considerations related to the collection of personally identifiable and potentially health-relevant data, and propose recommendations for the responsible and safe use of the system.

\subsection{Collection of (Sensitive) Video \& Health Data}
Developing and integrating methods that automatically collect personal identifiable and potentially health relevant information comes with additional responsibilities. While informing participants and clinicians on individual social interaction behaviors can be beneficial for diagnostic purposes and planning targeted social skills trainings, video data might be susceptible to misuse, including deceptive purposes like surveillance, identity theft, and personality profiling. Companies might base hiring decisions or performance evaluations on employees’ ``social interaction skills" in certain circumstances and malicious agents might make use of potentially vulnerable people (i.e., socially anxious) in potentially vulnerable contexts (i.e., high acute stress levels) for their own purposes (i.e., selling food, placing advertisements). It is important to inform potential users but also the general population on both the potential benefits and risks associated with these methods.

\subsection{Recommendations for Responsible Deployment}
The aforementioned harmful uses clearly cross legal boundaries and are partially addressed in national and international laws, such as the General Data Protection Regulation (GDPR) and the Health Insurance Portability and Accountability Act (HIPAA). However, these risks should also be minimized within the research community by taking proactive steps toward data minimization, security, and transparency. We recommend that the Kids-SIT should always be integrated in a study or diagnostic context where informed consent, debriefing, cancellation and data deletion possibility at any time, and contact to the researchers are made sure. We see ethical approval but also consultation with the respective data protection as mandatory in all cases. To safeguard sensitive data from unauthorized access, breaches, and misuse, encryption protocols, access controls, and secure storage solutions should be implemented before data collection.

\bibliographystyle{IEEEtran}
\bibliography{egbib}

\end{document}